\documentclass[a4paper,USenglish]{lipics-v2018}

\usepackage{microtype}

\usepackage[title]{appendix}
\usepackage{amsthm}
\usepackage{longtable}
\usepackage{tikz}
\usetikzlibrary{calc}
\usepackage{enumerate}
\usepackage{enumitem}

\theoremstyle{definition}
\newtheorem*{mydefinition}{Definition}

\title{CASPaxos: Replicated State Machines without logs}

\titlerunning{CASPaxos}

\author{Denis Rystsov}{Microsoft}{derystso@microsoft.com}{}{}

\authorrunning{Denis Rystsov}

\Copyright{Denis Rystsov}

\subjclass{Information systems $\rightarrow$ Distributed storage}


\supplement{https://github.com/gryadka/js}

\acknowledgements{We thank Tobias Schottdorf and Greg Rogers for writing TLA+ specs for CASPaxos, and Ezra Hoch, Peter Bourgon, Lisa Kosiachenko, Reuben Bond and Murat Demirbas for valuable comments and suggestions.}

\EventEditors{John Q. Open and Joan R. Access}
\EventNoEds{2}
\EventLongTitle{42nd Conference on Very Important Topics (CVIT 2016)}
\EventShortTitle{CVIT 2016}
\EventAcronym{CVIT}
\EventYear{2016}
\EventDate{December 24--27, 2016}
\EventLocation{Little Whinging, United Kingdom}
\EventLogo{}
\SeriesVolume{42}
\ArticleNo{23}
\nolinenumbers 
\hideLIPIcs  

\begin{document}
    
    \maketitle
    
    \begin{abstract}
        CASPaxos is a wait-free, linearizable, multi-writer multi-reader register in unreliable, asynchronous networks supporting arbitrary update operations including compare-and-set (CAS). The register acts as a replicated state machine providing an interface for changing its value by applying an arbitrary user-provided function (a command). Unlike Multi-Paxos and Raft which replicate the log of commands, CASPaxos replicates state, thus avoiding associated complexity, reducing write amplification, increasing concurrency of disk operations and hardware utilization.
    
        The paper describes CASPaxos, proves its safety properties and evaluates the characteristics of a CASPaxos-based prototype of key-value storage.
    \end{abstract}

\section{Introduction}

    Replicated state machine (RSM) protocols allow a collection of nodes to work as a state machine tolerating non-byzantine node failures and communication problems. The protocols guarantee safety in the presence of arbitrary node crashes, message loss, and out-of-order delivery; and preserve liveness when at most $\lfloor \frac{N-1}2 \rfloor$ of $N$ machines are down or disconnected.

    RSMs and fault-tolerant strongly consistent (linearizable) storages are equivalent in a sense that one can be implemented via the other, and vice versa. This explains why Multi-Paxos\cite{lamport01} and Raft\cite{raft} are widely used in the industry as a foundation of distributed systems (e.g. Chubby\cite{chubby}, Etcd\footnote{\href{https://github.com/coreos/etcd}{https://github.com/coreos/etcd}}, Spanner\cite{spanner}).

    Despite the wide adoption, there are a lot of indications that those protocols are complex. Diego Ongaro and John Ousterhout write in "In Search of an Understandable Consensus Algorithm"\cite{raft}:

    \begin{quote}
        In an informal survey of attendees at NSDI 2012, we found few people who were comfortable with Paxos, even among seasoned researchers. We struggled with Paxos ourselves; we were not able to understand the complete protocol until after reading several simplified explanations and designing our own alternative protocol, a process that took almost a year
    \end{quote}

    Google's engineers write about their experience of building a Paxos-based database in the "Paxos Made Live"\cite{chubby} paper:

    \begin{quote}
        Despite the existing literature in the field, building such a database proved to be non-trivial \ldots{} While Paxos can be described with a page of pseudo-code, our complete implementation contains several thousand lines of C++ code \ldots{} There are significant gaps between the description of the Paxos algorithm and the needs of a real-world system.
    \end{quote}

    Complexity of the RSM protocols may explain challenges in implementations. Kyle Kingsbury made a comprehensive research\footnote{\href{https://aphyr.com/tags/jepsen}{https://aphyr.com/tags/jepsen}} of the distributed consistent storages and found violations of linearizability in some version of almost every system he tested including MongoDB, Etcd, Consul, RethinkDB, VoltDB, and CockroachDB.

    Besides complexity, those protocols are subject to temporary loss of availability. The "There Is More Consensus in Egalitarian Parliaments" paper\cite{epaxos} describes the negative implications of a leader-based system which are applicable both to Multi-Paxos and Raft:

    \begin{quote}
        Traditional Paxos variants are sensitive to both long-term and transient load spikes and network delays that increase latency at the master \ldots{} this single-master optimization can harm availability: if the master fails, the system cannot service requests until a new master is elected \ldots{} Multi-Paxos has high latency because the local replica must forward all commands to the stable leader.
    \end{quote}

    {\bf Contributions.} We present CASPaxos, a novel leaderless protocol for building RSM that avoids complexity associated with Multi-Paxos/Raft and availability implementations of having a leader and reduces write amplification.

    Multi-Paxos based system is a RSM built on top of a replicated log which treats every log entry as a command. The replicated log is composed of an array of Synod\cite{lamport01} (also known as Single Decree Paxos) instances. According to the Raft paper, its complexity comes from the composition rules:

    \begin{quote}
        We hypothesize that Paxos’ opaqueness derives from its choice of the single-decree subset as its foundation \ldots{} The composition rules for Multi-Paxos add significant additional complexity and subtlety.

        One reason is that there is no widely agreed upon algorithm for multi-Paxos. Lamport’s descriptions are mostly about single-decree Paxos; he sketched possible approaches to multi-Paxos, but many details are missing. As a result, practical systems bear little resemblance to Paxos. Each implementation begins with Paxos, discovers the difficulties in implementing it, and then develops a significantly different architecture \ldots{} real implementations are so different from Paxos that the proofs have little value
    \end{quote}

    The main idea of CASPaxos is to replicate state instead of the log of commands. We achieve it by extending Synod instead of using it as a building block. As a consequence, there is no composition and the associated complexity. It allows our implementation\footnote{\href{https://github.com/gryadka/js}{https://github.com/gryadka/js}} to be less than 500 lines of code.

    Being just an extension of Synod, CASPaxos uses its symmetric peer-to-peer approach and automatically achieves the goals set in the EPaxos\cite{epaxos} paper: (1) optimal commit latency in the wide-area when tolerating failures under realistic conditions; (2) uniform load balancing across all replicas (thus achieving high throughput); and (3) graceful performance degradation when replicas are slow or crash.

    {\bf Correctness}. The protocol has a formal proof included in the appendix \ref{appendix:proof}. Also Tobias Schottdorf and Greg Rogers independently model checked the protocol with TLA+\footnote{\href{https://tschottdorf.github.io/single-decree-paxos-tla-compare-and-swap}{https://tschottdorf.github.io/single-decree-paxos-tla-compare-and-swap}, \href{https://medium.com/@grogepodge/tla-specification-for-gryadka-c80cd625944e}{https://medium.com/@grogepodge/tla-specification-for-gryadka-c80cd625944e}}.

\section{Protocols}

    We begin by briefly describing the Synod protocol from the perspective of master-master replication, followed by a step by step comparison with CASPaxos.

    \begin{figure}[!h]
        \centering
        \begin{tikzpicture}[y=-1cm]
            \node at (0,-0.5)[scale=0.8]{Client};
            \node at (2,-0.5)[scale=0.8]{Proposer};
            \node at (4.5,-0.5)[scale=0.8]{Acceptor A};
            \node at (6.5,-0.5)[scale=0.8]{Acceptor B};
            \node at (8.5,-0.5)[scale=0.8]{Acceptor C};
          
            \draw (0,0) -- (0,5.6);
            \draw (2,0) -- (2,5.6);
            \draw (4.5,0) -- (4.5,5.6);
            \draw (6.5,0) -- (6.5,2.05);
            \draw (8.5,0) -- (8.5,5.6);
            \draw (6.4,2.05) -- (6.6,2.05);
        
            \draw[dotted] (-0.2,0.75) -- (8.6,0.75);
            \draw[dotted] (-0.2,2.55) -- (8.6,2.55);
            \draw[dotted] (-0.2,4.05) -- (8.6,4.05);
        
            \begin{scope}[very thick]
                \draw[->] (0,0.5) -- (2,0.5) node[above, midway, scale=0.8]{set 3};
            
                \node at (1,1.6)[scale=0.8]{Propose};
                \draw[->] (2,1) -- (4.5,1);
                \draw[->] (2,1.3) -- (6.5,1.3);
                \draw[->] (2,1.6) -- (8.5,1.6);
                \draw[<-, dashed] (2,1.9) -- (4.5,1.9);
                \draw[<-, dashed] (2,2.2) -- (8.5,2.2);
        
                \node at (1,3.35)[scale=0.8]{Accept};
                \draw[->] (2,2.9) -- (4.5,2.9);
                \draw[->] (2,3.2) -- (8.5,3.2);
                \draw[<-, dashed] (2,3.5) -- (4.5,3.5);
                \draw[<-, dashed] (2,3.8) -- (8.5,3.8);
        
                \draw[<-, dashed] (0,4.3) -- (2,4.3) node[below, midway, scale=0.8]{ok};
            \end{scope}
        \end{tikzpicture}
        \caption{Synod sequential diagram}
    \end{figure}
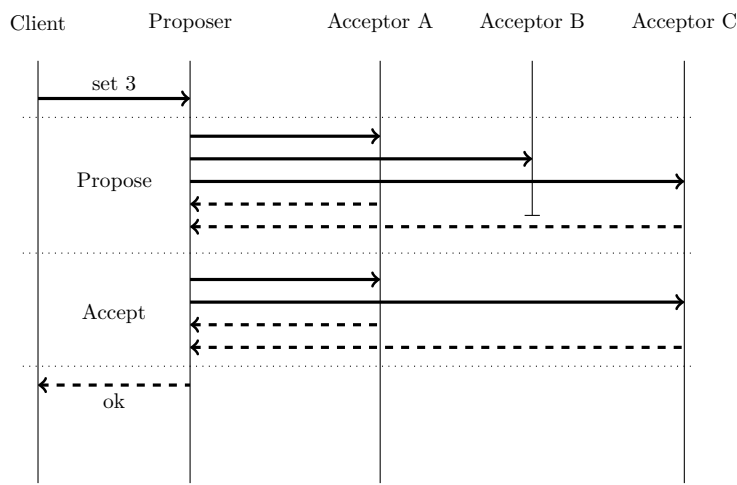
    
\subsection{Synod}
    
    An implementation of the Synod protocol is an initializable-only-once distributed register. When several clients try to initialize it concurrently - at most one client succeeds. Once a client receives a confirmation, all the follow-up initializations must return the already chosen value.
    
    The register belongs to the CP category of the CAP theorem\footnote{\href{https://en.wikipedia.org/wiki/CAP\_theorem}{https://en.wikipedia.org/wiki/CAP\_theorem}}, gives up availability when more than $\lfloor \frac{N-1}2 \rfloor$ nodes are down and always preserves consistency (linearizability).
    
    Each node in the system plays the role either of client, proposer or acceptor.
    
    \begin{description}[align=left]
        \item [Clients] initiate a request by communicating with a proposer; clients may be stateless, the system may have arbitrary numbers of clients.
        \item [Proposers] perform the initialization by communicating with acceptors. Proposers keep minimal state needed to generate unique increasing update IDs (ballot numbers), the system may have arbitrary numbers of proposers.
        \item [Acceptors] store the accepted value; the system should have $2F+1$ acceptors to tolerate $F$ failures.
    \end{description}
    
    It's convenient to use tuples as ballot numbers. To generate it a proposer combines its comparable ID with a local increasing counter: (counter, ID). To compare ballot tuples, we should compare the first component of the tuples and use ID only as a tiebreaker.
    
    When a proposer receives a conflicting message from an acceptor, it should update its counter to avoid a conflict in the future.

\subsection{CASPaxos}

    An implementation of CASPaxos is a rewritable distributed register. Clients change its value by submitting side-effect free functions which take the current state as an argument and yield new as a result. Out of the concurrent requests only one can succeed, once a client gets a confirmation of the change it's guaranteed that all future states are its descendants: there exists a chain of changes linking them together.

    Just like with Synod, it's a CP-system, and it requires $2F+1$ nodes to tolerate $F$ failures. Also, it uses the same roles: clients, proposers, and acceptors, and a very similar two-phase state transition mechanism.

    Let's review the Synod and CASPaxos protocols step-by-step.

    \begin{center}
        \begin{longtable}{p{15em}|p{15em}} 
            \hline
            {\bf Synod}
            &
            {\bf CASPaxos} \\ 
            \hline
            \endfirsthead
        
            \endhead
            \endfoot
            \endlastfoot
          
            A client proposes the $val_0$ value to a proposer.
            &
            A client submits the $f$ change function to a proposer. \\
          
            \hline
            
            The proposer generates a ballot number, $B$, and sends "prepare" messages containing   that number to the acceptors.
            &
            The proposer generates a ballot number, $B$, and sends "prepare" messages containing   that number to the acceptors. \\
            
            \hline
            
            {\bf An acceptor}
            &
            {\bf An acceptor} \\[6pt]
            
            
            Returns a conflict if it already saw a greater ballot number.
            &
            Returns a conflict if it already saw a greater ballot number.
            \\[6pt]
            
            
            Persists the ballot number as a promise and returns a confirmation either with an   empty value (if it hasn't accepted any value yet) or with a tuple of an accepted   value and its ballot number.
            &
            Persists the ballot number as a promise and returns a confirmation either with an   empty value (if it hasn't accepted any value yet) or with a tuple of an accepted   value and its ballot number.
            \\[6pt]
            
            \hline
            
            {\bf The proposer}
            &
            {\bf The proposer} \\[6pt]
            
            
            Waits for the $F+1$ confirmations
            &
            Waits for the $F+1$ confirmations \\[6pt]
            
            
            If they all contain the empty value, then the proposer defines the current state as   $val_0$ otherwise it picks the value of the tuple with the highest ballot number.
            &
            If they all contain the empty value, then the proposer defines the current state as   $\emptyset$ otherwise it picks the value of the tuple with the highest ballot number.
            \\[6pt]
            
            
            Sends the current state along with the generated ballot number $B$ (an "accept"   message) to the acceptors.
            &
            Applies the $f$ function to the current state and sends the result, new state, along   with the generated ballot number $B$ (an "accept" message) to the acceptors.
            \\[6pt]
            
            \hline
            
            {\bf An acceptor}
            &
            {\bf An acceptor} \\[6pt]
            
            
            Returns a conflict if it already saw a greater ballot number.
            &
            Returns a conflict if it already saw a greater ballot number.
            \\[6pt]
            
            
            Erases the promise, marks the received tuple (ballot number, value) as the accepted   value and returns a confirmation
            &
            Erases the promise, marks the received tuple (ballot number, value) as the accepted   value and returns a confirmation
            \\[6pt]
            
            \hline
            
            {\bf The proposer}
            &
            {\bf The proposer} \\[6pt]
            
            
            Waits for the $F+1$ confirmations
            &
            Waits for the $F+1$ confirmations. \\[6pt]
            
            
            Returns the current state the client.
            &
            Returns the new state to the client. \\[6pt]
            
            \hline
        \end{longtable}
    \end{center}

    As we see, the CASPaxos's state transition is almost identical to the Synod's initialization, and if we use

        $$x \to \mbox{if}\; x = \emptyset \;\mbox{then}\; val_0\; \mbox{else}\; x$$
    
    as the change function then it indeed becomes identical.

    By choosing a different set of change functions, we can turn CASPaxos into a distributed register and atomically change its value with arbitrary user-defined function in one operation. For example, the following change functions implement a register supporting CAS.

    \begin{itemize}
        \item To {\bf initialize} a register with $val_0$ value
        $$x \to \mbox{if}\; x = \emptyset \;\mbox{then}\; (0, val_0)\; \mbox{else}\; x$$
        
        \item To {\bf update} a register to value $val_1$ if the current version is $5$
        $$x \to \mbox{if}\; x = (5, \ast) \;\mbox{then}\; (6, val_1)\; \mbox{else}\; x$$
        
        \item To {\bf read} a value
        $$x \to x$$
    \end{itemize}

    With this specialization, the protocol is very similar to Bizur\cite{bizur}.

\subsubsection{One-round trip optimization}\label{1rtt}

    Since the prepare phase doesn't depend on the change function, the next prepare message can piggyback on the current accept message to reduce the number of round trips from two to one.

    In this case, a proposer caches the last written value, and the clients should use that proposer to initiate the state transition. The optimization doesn't compromise safety if a client sends a request to another proposer.

\subsubsection{Low-latency and high-throughput consensus across WAN deployments}

    WPaxos\cite{wpaxos} paper describes how to achieve low-latency and high-throughput consensus across wide area network through object stealing. It leverages the flexible quorums\cite{fpaxos} idea to reduce WAN communication costs. Since CASPaxos is an extension of Synod and supports FPaxos (see the proof in the appendix \ref{appendix:fpaxos}), it supports the WPaxos optimization too.

\subsection{Cluster membership change}

    Cluster membership change is a process of changing the set of nodes executing a distributed system without violating safely and liveness properties. It's crucial to have this process because it solves two problems:

    \begin{enumerate}
    \item {\it How to adjust fault tolerance properties of a system}. With time the fault tolerant requirements may change. Since a CASPaxos-based system of size $N$ tolerates up to $\lfloor \frac{N-1}2 \rfloor$ crashes, a way to increase/decrease size of a cluster is also a way to increase/decrease resiliency of the system.

    \item {\it How to replace permanently failed nodes.} CASPaxos tolerates transient failures, but the hardware tends to break, so without a replacement eventually, more than $\lfloor \frac{N-1}2 \rfloor$ nodes crash, and the system becomes unavailable. A replacement of a failed node in the $N$ nodes cluster can be modeled as a shrinkage followed by an expansion.
    \end{enumerate}

    The process of membership change is based on Raft's idea of joint consensus where two different configurations overlap during transitions. It allows the cluster to continue operating normally during the configuration changes.

    The proof of applicability of this idea to CASPaxos is based on two observations:

    \begin{itemize}
        \item {\bf Flexible quorums} It has been observed before that in a Paxos-based system the only requirement for the "prepare" and "accept" quorums is the intersection \cite{abcds}\cite{vertical}\cite{fpaxos}. For example, if the cluster size is $4$, then we may require $2$ confirmations during the "prepare" phase and $3$ during the "accept" phase.
    
        \item {\bf Network equivalence} If a change in the behavior of a Paxos-based system can be explained by delaying or omitting the messages between the nodes, then the change doesn't affect consistency because Paxos tolerates the interventions of such kind. It gives freedom in changing the system as long as the change can be modeled as a message filter on top of the unmodified system.
    \end{itemize}

\subsubsection{Expansion of a cluster with an odd number of nodes}

    The protocol for changing the set of acceptors from $A_1 \cdots A_{2F+1}$ to $A_1 \cdots A_{2F+2}$:
    \begin{enumerate}
        \item Turn on the $A_{2F+2}$ acceptor.
        
        \item Connect to each proposer and update its configuration to send the "accept" messages to the $A_1 \cdots A_{2F+2}$ set of acceptors and to require $F+2$ confirmations during the "accept" phase.\label{dual}
        
        \item Pick any proposer and execute the identity state transition function $x \to x$.\label{rescan}
        
        \item Connect to each proposer and update its configuration to send "prepare" messages to the $A_1 \cdots A_{2F+2}$ set of acceptors and to require $F+2$ confirmations.
    \end{enumerate}

    From the perspective of the $2F+1$ nodes cluster, the second step can be explained with the network equivalence principle, so the system keeps being correct. When all proposers are modified the system also works as a $2F+2$ nodes cluster with $F+1$ "prepare" quorum and $F+2$ "accept" quorum.

    After the read operation finishes the state of the cluster becomes valid from the $F+2$ perspective, so we can forget about the $F+1$ interpretation. The last step switches the system from the reduced "prepare" quorum to the regular.

    The same sequence executed in the reverse order shrinks cluster with an even number of nodes.

\subsubsection{Expansion of a cluster with an even number of nodes}

    The $A_1 \cdots A_{2F+2}$ to $A_1 \cdots A_{2F+3}$ extension protocol is more straightforward because we can treat a $2F+2$ nodes cluster as a $2F+3$ nodes cluster where one node had been down from the beginning:
    \begin{enumerate}
        \item Connect to each proposer and update its configuration to send the prepare \& accept messages to the $A_1 \cdots A_{2f+3}$ set of acceptors.

        \item Turn on the $A_{2f+3}$ acceptor.
    \end{enumerate}

    It's important to notice that the procedure works based on the assumption that the $2f+3^{\mbox{th}}$ node has always been down. If a cluster gets into even configuration from an odd configuration, then it's necessary to execute identity state transition (re-scan) before the extension to avoid data loss. 

    Otherwise, it's possible to sequentially replace every acceptor with an empty acceptor, lose all data and violate linearizability.

\subsubsection{Optimization}

    In a key-value storage implemented as an array of independent labeled CASPaxos instances, we need to perform the \ref{rescan} step for each instance (key). It results in a rescan of all record. If the storage has $K$ keys then during the $A_1 \cdots A_{2F+1}$ to $A_1 \cdots A_{2F+2}$ transition the rescan moves $K(2F+3)$ records.

    The goal of the identity state transition is to make the state of the cluster valid from the $F+2$ perspective. The alternative way to reach this state is to replicate a majority of $A_1 \cdots A_{2F+1}$ nodes for any moment after the \ref{dual} step into $A_{2F+2}$ resolving conflicts by choosing an accepted value with higher ballot number. Thus reducing the rescan cost from $K(2F+3)$ to $K(F+1)$.

    A background catch-up process keeping acceptors in sync up to some recent moment may reduce the cost further to $(K-k) + k(F+1)$ where $k$ is the number of updated keys since the last moment of sync.

\subsubsection{Changing the number of proposers}

    Consistency and availability properties don't depend on the number of proposers so that they can be added and removed at any time. The only caveat is the procedures like shrinkage-expansion of acceptors and deletion (\ref{deletion}) which need to update all proposers as one of the steps. Fortunately, those steps are idempotent so we can bring a proposer down, update the list of all proposers and on the next attempt the steps succeed.

    An algorithm to remove a proposer:

    \begin{enumerate}
    \item Turn off the proposer.
    \item Update the list of proposers of the GC process.
    \item Update the list of proposers of the process controlling the shrinkage-expansion of acceptors.
    \end{enumerate}

    An algorithm to add a proposer:

    \begin{enumerate}
    \item Update the list of proposers of the process controlling the shrinkage-expansion of acceptors.
    \item Update the list of proposers of the GC process.
    \item Turn on the proposer.
    \end{enumerate}

\section{A CASPaxos-based key-value storage}

    The lightweight nature of CASPaxos creates new ways for designing distributed systems with complex behavior. In this section, we'll discuss a CASPaxos-based design for a key-value storage and compare a research prototype with Etcd, MongoDB and other distributed databases.

    Instead of designing a key-value storage as a single RSM we represent it as a set of labeled CASPaxos instances. Gryadka\footnote{\href{https://github.com/gryadka/js}{https://github.com/gryadka/js}} is a prototype implementing this idea.

\subsection{How to delete a record}\label{deletion}

    CASPaxos supports only update (change) operation so to delete a value a client should update a register with an empty value (a tombstone). The downside of this approach is the space inefficiency: even when the value is empty, the system still spends space to maintain information about the removed register.

    The straightforward removal of this information may introduce consistency issues. Consider the following state of the acceptors.

    \begin{figure}[!h]
        \centering
        \begin{tabular}{ r|r|r|r }
            & Promise & Ballot & State \\ \hline
            Acceptor A && 2 & 42 \\
            Acceptor B && 3 & $\emptyset$ \\
            Acceptor C && 3 & $\emptyset$ \\
        \end{tabular}
    \end{figure}

    According to the CASPaxos protocol, a read operation (implemented as $x \to x$ change function) should return $\emptyset$. However, if we decide to remove all information associated with the register and the read request hits the system during the process when the data on acceptors B and C have already gone then the outcome is $42$ which violates linearizability.

    An increasing of the "accept" quorum to $2F+1$ on writing an empty value before the removal solves the problem, but it makes the system less available since it's impossible to remove a register when at least one node is down.

    A multi-step removal process fixes this problem.

    \begin{enumerate}
        \item On a delete request, a proposer writes a tombstone with regular $F+1$ "accept" quorum, schedules a garbage collection operation and confirms the request to a client.

        \item The garbage collection operation (in the background):\label{GC}

        \begin{enumerate}
            \item Replicates an empty value to all nodes by executing the identity transform with max quorum size ($2F+1$).\label{tombstone}

            \item Connects to each proposer, invalidates its cache associated with the removing key (see one-round trip optimization \ref{1rtt}), fast-forwards its counter to be greater than the tombstone's ballot number and increments proposer's age.

            \item Connects to each acceptor and asks it to reject messages from proposers if their age is younger than the corresponding age from the previous step.

            \item Removes the register from each acceptor if it contains the tombstone from the \ref{tombstone} step.
        \end{enumerate}
    \end{enumerate}

    Each step of the GC process is idempotent so if any acceptor or proposer is down the process reschedules itself.

    Invalidation of the proposer's caches and the age check are necessary to eliminate the lost delete anomaly, a situation when a message delayed by a channel (or an accept message corresponding to a change of the cached value) revives a value without a causal link to the deletion event.

    The update of the counters is necessary to avoid the lost update anomaly which may happen when a concurrently updated value has lesser ballot number than the tombstone's ballot number, and a reader prioritizes the tombstone over the new value.

    To make the age check possible, proposers should include their age into every message they send, and acceptors should persist age per proposer set by GC process.

\subsection{Low latency}

    The following properties of CASPaxos help achieve low latency:

    \begin{itemize}[noitemsep]
      \item It isn't a leader-based protocol so a proposer should not forward all requests to a specific node to start executing them.

      \item Requests affecting different key-value pairs do not interfere.

      \item It uses 1RTT when the requests affecting the same key land on the same proposer.

      \item No acceptor is special, so a proposer ignores slow acceptors and proceeds as soon as quorum is reached.

      \item An ability to use user-defined functions as state transition functions reduces two steps transition process (read, modify, write) into one step process.
    \end{itemize}
    
    We compared Gryadka with Etcd and MongoDB to check it. All storages were tested in the same environment. Gryadka, Etcd, and MongoDB were using three DS4\_V2 nodes configuration deployed over WAN in the Azure's\footnote{\href{https://azure.microsoft.com}{https://azure.microsoft.com}} datacenters in the "West US 2", "West Central US" and "Southeast Asia" regions.
    
    Each node has a colocated client which in one thread in a loop was reading a value, incrementing and writing it back. All clients used their keys to avoid collisions. During the experiment latency (average duration of read-modify-write operation) was measured per each client (region).
    
    \begin{figure}[!htb]
        \centering
        \begin{tabular}{c|r|r|r|}
            \cline{2-4}
            & \multicolumn{3}{|c|}{Latency} \\
            \cline{2-4}
            & MongoDB (3.6.1) & Etcd (3.2.13) & Gryadka (1.61.8) \\
            \hline
            \multicolumn{1}{|l|}{West US 2} & 1086 ms & 679 ms & 47 ms \\
            \hline
            \multicolumn{1}{|l|}{West Central US} & 1168 ms & 718 ms & 47 ms \\
            \hline
            \multicolumn{1}{|l|}{Southeast Asia} & 739 ms & 339 ms & 356 ms \\
            \hline
        \end{tabular}
    \end{figure}
    
    The result matches our expectation especially if we take into account delay between datacenters and the leader/leaderless nature of MongoDB, Etcd, and Gryadka.
    
    \begin{figure}[!h]
        \centering
        \begin{tabular}{llr|}
            \cline{3-3}
            & & \multicolumn{1}{|l|}{RTT} \\
            \hline
            \multicolumn{1}{|l|}{West US 2} & \multicolumn{1}{|l|}{West Central US} & 21.8 ms\\
            \hline
            \multicolumn{1}{|l|}{West US 2} & \multicolumn{1}{|l|}{Southeast Asia} & 169 ms\\
            \hline
            \multicolumn{1}{|l|}{West Central US} & \multicolumn{1}{|l|}{Southeast Asia} & 189.2 ms\\
            \hline
        \end{tabular}
    \end{figure}
    
    The leaders of MongoDB and Etcd were in the "Southeast Asia" region so to execute an operation the "West US 2" node needs additional round trip to forward a request to the leader and to receive a response (169 ms). Then the leader needs to write the change to the majority of nodes and get confirmations (0 ms and 169 ms). Since the iteration consists of reading and writing operations, in total "West US 2" node requires at least $676 \mbox{ms} = 2 \cdot (169 \mbox{ms} + 169 \mbox{ms})$. For the "West Central US" node the estimated latency is $716.4 \mbox{ms} = 2 \cdot (169 \mbox{ms} + 189.2 \mbox{ms})$, for "Southeast Asia" it's $338 \mbox{ms} = 2 \cdot 169 \mbox{ms}$.
    
    Gryadka doesn't forward requests so the corresponding estimated latencies are $43.6 \mbox{ms} = 2 \cdot 21.8 \mbox{ms}$, $43.6 \mbox{ms} = 2 \cdot 21.8 \mbox{ms}$ and $338 \mbox{ms} = 2 \cdot 169 \mbox{ms}$.
    
    Network fluctuations and storage implementation details may explain the minor difference between estimated and measured latencies.
    
    As we see, the underlying consensus protocol plays an essential role in the performance of the system.

\subsection{Fault-tolerance}

    The EPaxos paper explains how the leader-based consensus protocols lead to cluster-wide unavailability when a leader crashes or is isolated from the cluster:

    \begin{quote}
    With Multi-Paxos, or any variant that relies on a stable leader, a leader failure prevents the system from processing client requests until a new leader is elected. Although clients could direct their requests to another replica (after they time out), a replica will usually not try to become the new leader immediately
    \end{quote}

    CASPaxos doesn't suffer from this behavior because all of its nodes of the same role are homogeneous so when any of them is isolated it doesn't affect processes running on the other nodes. An experimental study\footnote{\href{https://github.com/rystsov/perseus}{https://github.com/rystsov/perseus}} of distributed consistent databases with default settings during a leader isolation accident supports this a priori reasoning - all systems but Gryadka has a non-zero complete unavailability window (all client's operations halt).

    \begin{figure}[!h]
        \centering
        \begin{tabular}{|l|l|l|r|}
            \hline
            Database & Version & Protocol & Unavailability\\
            \hline
            \hline
            Gryadka & 1.61.8 & CASPaxos & $<$ 1s\\
            \hline
            CockroachDB & 1.1.3 & MultiRaft & 7s\\
            Consul & 1.0.2 & Raft & 14s\\
            Etcd & 3.2.13 & Raft & 1s\\
            RethinkDB & 2.3.6 & Raft & 17s\\
            Riak & 2.2.3 & Vertical Paxos & 8s\\
            TiDB & 1.1.0 & MultiRaft & 15s\\
            \hline
        \end{tabular}
    \end{figure}

\section{Comparison with Related Work}

    {\bf Consensus.} CASPaxos and Raft/Multi-Paxos have different trade-offs. CASPaxos replicates state on each change request so it's impractical for RSM with a heavy monolithic state. The absence of logs eliminates artificial synchronization and reduces write amplification. Independent registers allow batching and out of order writing achieving high concurrency of disk operations thus better hardware utilization. Besides that, the absence of leader increases availability and reduces latency.

    {\bf Registers.} Protocols of atomic distributed registers, described in "Sharing memory robustly in message-passing systems"\cite{abd} and "On the Efficiency of Atomic Multi-reader, Multi-writer Distributed Memory"\cite{mwmr} papers, has similar structure to CASPaxos. But they don't provide conditional writes or other concurrency control primitives which makes them impractical for concurrent environments and makes it's imposible to support client-side transactions.

\section{Conclusion}

    We demonstrated that CASPaxos is a simple RSM protocol without availability implications of having a leader with high concurrency of disk operations and hardware utilization.

    The possible applications of the protocol are key-value storages and stateful actor based services such as Microsoft Orleans\footnote{\href{https://dotnet.github.io/orleans}{https://dotnet.github.io/orleans}}. The indirect contributions of our work is the proof of safety properties which is also applicable to the other variants of Paxos.

    \newpage

    \bibliography{caspaxos}

    \newpage

    \begin{appendices}
        \section{Proof}
        \label{appendix:proof}

        We want to prove that for any two acknowledged changes one is always a descendant of another.
        
        We'll do it by reasoning about the four types of the events:
        
        \begin{itemize}
            \item $e \in \ddot{E}^1$ - an acceptor received a "prepare" message and replied with "promise"
            \item $e \in \bar{E}^1$ - a proposer sent an "accept" message
            \item $e \in \ddot{E}^2$ - an acceptor accepted a new value
            \item $e \in \bar{E}^2$ - a proposer received a majority of confirmations
        \end{itemize}
        
        Let's review the structure of the events.
        
        \begin{figure}[!h]
            \centering
            \begin{minipage}{0.47\textwidth}
                \centering
                \begin{tabular}{|p{1cm}|p{4cm}|}
                    \hline
                    \multicolumn{2}{|c|}{$e \in \ddot{E}^1$ - promised event}\\
                    \hline
                    $e.ts$ & acceptor's local time\\
                    \hline
                    $e.b$ & ballot number of current change round\\
                    \hline
                    $e.ret.b$ & a ballot number of the last accepted value\\
                    \hline
                    $e.ret.s$ & the last accepted value\\
                    \hline
                    $e.node$ & acceptor's id\\
                    \hline
                    \hline
                    $s(e)$ & is equal to $e.ret.s$\\
                    \hline
                \end{tabular}
            \end{minipage}
            \hspace{\fill} 
            \begin{minipage}{0.47\textwidth}
                \centering
                \begin{tabular}{|p{1cm}|p{4cm}|}
                    \hline
                    \multicolumn{2}{|c|}{$e \in \bar{E}^1$}\\
                    \hline
                    $e.ts$ & proposer's local time\\
                    \hline
                    $e.b$ & ballot number of current change round\\
                    \hline
                    $e.s$ & new value, result of a change function applied to a value corresponding to a promise with maximum ballot number\\
                    \hline
                    \hline
                    $s(e)$ & is equal to $e.s$\\
                    \hline
                \end{tabular}
            \end{minipage}
            
            \vspace*{1cm}
        
            \begin{minipage}{0.47\textwidth}
                \centering
                \begin{tabular}{|p{1cm}|p{4cm}|}
                    \hline
                    \multicolumn{2}{|c|}{$e \in \ddot{E}^2$ - accepted event}\\
                    \hline
                    $e.ts$ & acceptor's local time\\
                    \hline
                    $e.b$ & ballot number of current change round\\
                    \hline
                    $e.s$ & accepted new value\\
                    \hline
                    $e.r$ & $\{ x | x \in \ddot{E}^1 \land x.b = e.b \}$\\
                    \hline
                    $e.node$ & acceptor's id\\
                    \hline
                    \hline
                    $s(e)$ & is equal to $e.s$\\
                    \hline
                \end{tabular}
            \end{minipage}
            \hspace{\fill} 
            \begin{minipage}{0.47\textwidth}
                \centering
                \begin{tabular}{|p{1cm}|p{4cm}|}
                    \hline
                    \multicolumn{2}{|c|}{$e \in \bar{E}^2$ - acknowledged event}\\
                    \hline
                    $e.ts$ & proposer's local time\\
                    \hline
                    $e.b$ & ballot number of current change round\\
                    \hline
                    $e.s$ & accepted new value\\
                    \hline
                    $e.w$ & $\{ x | x \in \ddot{E}^2 \land x.b = e.b \}$\\
                    \hline
                    $e.node$ & proposer's id\\
                    \hline
                    \hline
                    $s(e)$ & is equal to $e.s$\\
                    \hline
                \end{tabular}
            \end{minipage}
        
            \caption{Structure of the events}
        \end{figure}
        
        We want to demonstrate that
        
        \begin{equation}
            \forall x,y \in \bar{E}^2 \;:\; x \to y \lor y \to x
        \end{equation}
        
        where $\to$ represents the "is a descendant" relation.
        
        \theoremstyle{definition}
        \begin{mydefinition}{\bf("is a descendant" relation)}
            What does it mean that one event is a descendant of another? Informally it means that there is a chain of state transitions leading from the state acknowledged in the initial event to the state acknowledged in the final event. Let's formalize it. We start by defining this relation on $\ddot{E}^2$ and then expand it to $\bar{E}^2$.
        
            By definition of CASPaxos, any accepted state is a function of previously accepted state, so
        
            \begin{equation} \label{eq:chain}
                \forall x \in \ddot{E}^2 \quad \exists ! f \quad \exists y \in \ddot{E}^2 : s(x) = f(s(y))
            \end{equation}
        
            When \ref{eq:chain} holds for $x$ and $y$ we write $y \sim x$ and $y = I^{-1}(x)$. Now we can define "is a descendant" relation on $\ddot{E}^2$ events as:
        
            \begin{equation}
                \forall x \in \ddot{E}^2 \; \forall y \in \ddot{E}^2 \;:\; x \to y \equiv x \sim y \lor (\exists z \in \ddot{E}^2 \;:\; x \to z \land z \to y)
            \end{equation}
          
            We can use that $\forall x \in \bar{E}^2 \; \forall y \in x.w \;:\; s(x) = s(y)$ is true (by definition) and continue "is a descendant" relation on $\bar{E}^2$:
          
            \begin{equation}
                \forall x \in \bar{E}^2 \; \forall y \in \bar{E}^2 \;:\; x \to y \equiv (\forall a \in x.w \; \forall b \in y.w \; a \to b)
            \end{equation}
        \end{mydefinition}
        
        \begin{lemma}
            The following statement proves that any two acknowledged changes one is always a descendant of another
          
            \begin{equation} \label{eq:step}
                \forall x \in \bar{E}^2 \; \forall y \in \ddot{E}^2 \;:\; x.b < y.b \implies x.b \leq I^{-1}(y).b
            \end{equation}
        \end{lemma}
        
        \begin{proof}
            Let $z_0 := y$ and $z_{n+1} := I^{-1}(z_{n})$. By definition, ballot numbers only increase: $z_{n+1}.b < z_{n}.b$, so we can use mathematical induction and \ref{eq:step} guarantees that $\exists k \;:\; z_k.b = x.b$ meaning $s(z_k) = s(x)$. Since $z_{k+1} \sim z_k$ we proved the following statement:
        
            \begin{equation} \label{eq:bd}
                \forall x \in \bar{E}^2 \; \forall y \in \ddot{E}^2 \;:\; x.b < y.b \implies x \to y
            \end{equation}
            
            Since $\forall y \in \bar{E}^2 \; \forall z \in y.w \;:\; y.b=z.b \land s(y)=s(z)$ then \ref{eq:bd} implies
            
            \begin{equation}
                \forall x \in \bar{E}^2 \; \forall y \in \bar{E}^2 \;:\; x.b < y.b \implies x \to y
            \end{equation}
            
            By definition, $\forall x \in \bar{E}^2 \; \forall y \in \bar{E}^2 \;:\; x.b < y.b \lor y.b < x.b$ so the latter means
            
            \begin{equation}
                \forall x \in \bar{E}^2 \; \forall y \in \bar{E}^2 \;:\; x \to y \lor y \to x
            \end{equation}
        \end{proof}
        
        \begin{theorem} \label{th:proof2}
            $$\forall x \in \bar{E}^2 \; \forall y \in \ddot{E}^2 \;:\; x.b < y.b \implies x.b \leq I^{-1}(y).b$$
        \end{theorem}
        
        \begin{proof}
            Let
            
            $$N = \{z.node \;|\; z \in x.w\} \cap \{z.node \;|\; z \in y.r\}$$
          
            $N$ isn't empty because "prepare" quorum intersects with "accept" quorum. Let $n \in N$, $u \equiv y.r |_n$ is a promised event happened on node $n$ and $w \equiv x.w |_n$ is an accepted event. By definition, $y.b = u.b$, $w.b = x.b$ and $x.b < y.b$ so
          
            \begin{equation}
                w.b < u.b
            \end{equation}
        
            Since an acceptor doesn’t accept messages with lesser ballot numbers than they already saw, the latter means $w.ts < u.ts$
          
            Let $P \equiv \{ x \;|\; x \in \ddot{E}^1 \land x.node = n\}$ and $A \equiv \{ x \;|\; x \in \ddot{E}^2 \land x.node = n \}$. Then for each $x \in P$, $x.ret.b$ is the ballot number of the latest accepted state, formally it means that:
          
            \begin{equation} \label{eq:last}
                \forall k \in P \quad k.ret.b = \max \{ l.b \;|\;l \in A \land l.ts < k.ts \} \\
            \end{equation}
          
            Since $w.b < u.b$, $w \in A$ and $u \in P$
          
            \begin{equation}
                w \in \{ z \in A \;|\; z.ts < u.ts \}
            \end{equation}
          
            With combination with \ref{eq:last} it implies:
          
            \begin{equation} \label{eq:final}
                x.b = w.b \leq \max \{ z.b \in A \;|\; z.ts < u.ts \} = u.ret.b
            \end{equation}
          
            By definition, a proposer picks a value out of majority of promise confirmations corresponding to the maximum ballot number, so:
          
            \begin{equation}
                I^{-1}(y).b = \max \{ z.ret.b | z \in y.r \}
            \end{equation}
          
            Combining with \ref{eq:final} we get:
          
            \begin{multline}
                x.b = w.b \leq \max \{ z \in A, z.ts < u.ts \} = \\
                = u.ret.b \leq \max \{ z.ret.b | z \in y.r \} = I^{-1}(y).b
            \end{multline}
          
            Which proves $x.b \leq I^{-1}(y).b$.
          
        \end{proof}
        
        \section{FPaxos}
        \label{appendix:fpaxos}

        The proof of CASPaxos \ref{appendix:proof} doesn't use the size of the promise/accept quorums and depends only on their intersection, so the same proof applies to FPaxos too.
            
    \end{appendices}

\end{document}